\documentclass[12pt]{article}
\usepackage{epsfig,graphicx, amssymb, amsmath, amsthm}

\begin{document}
\renewcommand{\textfraction}{0}

\theoremstyle{plain}
\newtheorem{theorem}{Theorem}
\newtheorem{lemma}{Lemma}
\numberwithin{equation}{section}
\allowdisplaybreaks

\title{Channel Model and Upper Bound on the Information Capacity of
the Fiber Optical Communication Channel Based on the Effects of
XPM Induced Nonlinearity}
\author{\normalsize
Hossein Kakavand \\
\small Department of Electrical Engineering \\[-5pt] \small Stanford University \\[-5pt]
\small Stanford, CA 94305\\[-5pt] \small
Hossein@stanford.edu}
\date{}
\maketitle \thispagestyle{empty}
\begin{abstract}
An upper bound to the information capacity of a
wavelength-division multiplexed optical fiber communication system
is derived in a model incorporating the nonlinear propagation
effects of cross-phase modulation (XPM). This work is based on the
paper by Mitra et al.~\cite{Mitra}, finding lower bounds to the
channel capacity, in which physical models for propagation are
used to calculate statistical properties of the conditional
probability distribution relating input and output in a single WDM
channel. In this paper we present a tractable channel model
incorporating the effects of cross phase modulation. Using this
model we find an upper bound to the information capacity of the
fiber optical communication channel at high SNR. The results
provide physical insight into the manner in which nonlinearities
degrade the information capacity.
\end{abstract}
\normalsize

\section{Introduction}
Communication via optical fibers has received a lot of attention
recently mainly due to the extremely high bandwidths and unique
propagation environments they provide. These characteristics provide
a rather appealing medium for multichannel communication through the
fiber, known as Wavelength Division multiplexing (WDM), which has
been the focus of much ongoing research and practical consideration,
recently~\cite{Mitra,Marcus,Killey}. However, optical fibers have a
number of nonlinear electromagnetic phenomena which affect the
propagation of signals. As we increase the rate of communication
through the fiber, the effect of these nonlinear phenomena will
limit the reliability of communication through the fiber. From an
information theoretic point of view these phenomena will limit the
information capacity of the fiber optical communication channel. In
order to derive this limit we need to study and model the effects of
such phenomena on the inputs to the channel as they propagate
through the fiber. This was first done by Mitra et al.~\cite{Mitra}.

In this paper we shall introduce the nonlinear effects in a fiber.
We shall model the effects of one of the nonlinear effects, known as
Cross Phase Modulation (XPM) on the propagation of signals through
the fiber and derive an upper bound on the capacity of the optical
channel considering the effects of XPM.

In section~\ref{sec:propagation} we shall describe the propagation
of an electromagnetic wave through the fiber, where our signal is
used to modulate the wave. In section~\ref{sec:ChannelModel} we
shall use our propagation equation to find a relation between the
input and output of a fiber, viewed as the communication channel.
This is done by considering the Green's function that relates the
input and output of the fiber medium. We shall then simplify the
Green's function in section~\ref{sec:simplify}. This simplification
results in a tractable channel model. We find an upper bound to the
information capacity of the simplified channel model in the high SNR
regime in section~\ref{sec:capacity}.

\section{Signal Propagation in Optical Fibers}
\label{sec:propagation} In this section we shall describe the
propagation of an input signal in an optical fiber, based on Mitra
et al.~\cite{Mitra} and Agrawal~\cite{Agrawal}, which shall be
used to derive a communication theoretic channel model in the next
sections.

In a single mode fiber used for Wavelength Division Multiplexed
(WDM) transmission, there are a number of channels $N$, each used
by an independent user. Each user uses an input signal $x_{k}(t)$
to modulate the carrier wave of frequency $\nu _{k}$ and has a
bandwidth of \(B \ll \nu_{k}\). We shall denote the carrier
frequency of the central channel by $\nu_0$, also the carrier
frequency spacing of neighboring channels is given by $\delta
\nu$. Since the users are independent the input signals to
different channels, $x_k(0,t)$, are also independent each having a
power constraint: $E_t(|x_k(0,t)|^2) \leq P,~~\forall k$. Along
the fiber $x_{k}(z,t)$ is the amplitude of the electric field
$\overline E_{k}$ in channel $k$ with propagation constant $\beta
_{k}$, were $z$ the distance travelled along the fiber. The total
electric field resulting from signals in N channels is given by,
\begin{eqnarray}
\overline E(z,t)=\sum_{k=-N/2}^{N/2}\left[x_{k}(z,t)
\exp(i(\beta_{k}z-2\pi\nu_{k}t))+x_{k}^{*}(z,t)
\exp(-i(\beta_{k}z-2\pi\nu_{k}t))\right]\exp(-\alpha z/2).\nonumber\\
\label{eqn:propagation1}
\end{eqnarray}
Equation~(\ref{eqn:propagation1}) accounts for the power loss due
to absorption in the fiber through the exponential decay factor
$\exp (-\alpha z/2)$. Also we take the direction of polarization
to be fixed and the traverse profile of the mode to be independent
of $z$. The propagation constants $\beta_{k}$, are frequency
dependent, making propagation through the fiber dispersive. This
phenomenon is known as Group Velocity Dispersion (GVD) which
results in different frequency components of a signal to travel at
different speeds, resulting in distortion of the input signal. It
also causes signals in different channels to travel at different
speeds.

In an optical fiber different frequency components interact
(couple) to generate new frequency components, this phenomenon is
known as Four Wave Mixing (FWM). Also due to the dependance of the
refractive index of the fiber on the power of propagating signals,
different channels interact. The interaction of a channel with
itself is known as Self Phase Modulation (SPM) and the interaction
of different channels with each other is known as Cross Phase
Modulation (XPM). For further details refer to~\cite{Mitra}
and~\cite{Agrawal}.

Given the phenomena explained above the signal in channel $k$,
$x_k(z,t)$ (written in the frame of reference that moves along the
fiber with the group velocity of the central channel) propagates
through the fiber based on the nonlinear Schrodinger equation,
\begin{equation}
[\partial_{z} + i \beta_{2}\partial_{t}^{2}] x_{k}(z,t) =
i\gamma[|x_{k}|^{2} + 2\sum_{l \neq k}|x_{l}|^{2}]\exp (-\alpha
z)x_{k}(z,t). \label{eqn:motion}
\end{equation}

Propagation along the fiber is thus characterized by a set of
$N+1$ coupled non-linear partial differential equations. In
Equation~(\ref{eqn:motion}), $\beta_k (\nu)$ has been Taylor
expanded about $\nu_0$ to take the form
\begin{equation*}
\beta_{k}(\nu)=\beta(\nu_{0}+k\delta\nu)= \beta_{0}+\beta_{1} k
\delta \nu + \beta_{2} k^{2} \delta \nu ^{2}/2 + O(\delta \nu ^{3}),
\end{equation*}
also, we have neglected the effect of Four Wave Mixing (FWM) since
it can be shown~\cite{Mitra} that the effect of FWM on the
information capacity can be studied separately. As will shall see,
SPM cannot be decoupled from the input signal and so its effect on
the capacity requires a different set of techniques, which can be
studied separately and will not be considered here~\cite{Mitra}.
However, XPM involves signals from other channels which are chosen
independently, and so for each signal the effect of other channels
is essentially random. Hence, in what follows we shall study the
effect of the stochastic effects of XPM on the information
capacity.

After the neglect of SPM and FWM, the effective channel model is
given by N+1 coupled nonlinear PDE's, known as the equations of
motion. The equation of motion for the central channel is given by,
\begin{eqnarray}
[\partial_{z}+\frac{\beta_{2}}{2} \partial_{t}^{2}]x_{0}(z,t)&=&
V_{0}(z,t)x_{0}(z,t),\label{eqn:channel}\\
V_{0}(z,t)&=&2 \gamma \sum_{l\neq0}|x_{l}(z,t)|^{2}\exp(-\alpha
z),\label{eqn:potential}
\end{eqnarray}

where $V_{0}(z,t)$ represents the XPM term for the central channel.
Equation~(\ref{eqn:channel}) is still non-linear because
$V_{0}(z,t)$ couples different channels to the central channel. Note
that, apart from the interchange of spatial and temporal
coordinates, $V_{0}(z,t)$ enters Equation~(\ref{eqn:channel}) in the
same way, and is therefore mathematically equivalent to, that of a
potential entering the Schrodinger equation~\cite{Feynman}.

\subsection{Equations of Motion}
The equations of motion are modeled by decoupling and linearizing
Equations (\ref{eqn:channel}) and (\ref{eqn:potential}). As
mentioned before the XPM term is mathematically equivalent to a
potential, $\nu(z,t)$. This potential is modeled as a Gaussian
random process, independent of all the channel inputs, in both
space and time with the same first and second order moments as the
XPM term, see Equations~(\ref{eqn:potenialfirst})
and~(\ref{eqn:potentialsecond}). This model follows from the
central limit theorem, considering the fact that at each instant
in time and space, $\sum_{l \neq 0}|x_{l}(z,t)|^{2}$ is the sum of
many independent random variables, each with a finite second
moment. Hence, the XPM term converges in distribution to a
Gaussian random process~\cite{Billingsley}. Note that while the
signal in each channel has both spatial and temporal correlation,
we are adding many such signals which travel at different speeds
in space, hence the correlations are lost. For further details
see~\cite{Mitra}.

Note that $\nu(z,t)$ does not depend on the signals in other
channels. The resulting model for propagation is linear, since the
equations for different channels have been decoupled. Hence, the
effective equation of motion for the central channel is given by,
\begin{equation}
[i\partial_{z}-\frac{\beta_{2}}{2}\partial_{t}^{2}]x_{0}(z,t)=\nu
(z,t) x_{0}(z,t) \label{eqn:propagation}
\end{equation}
\begin{equation}
E(\nu(z,t)) = E(V_0(z,t)),~~\forall t, \forall z,~~ 0 \leq z \leq
L \label{eqn:potenialfirst}
\end{equation}
\begin{equation}
E(\nu (z_1,t_1) \nu (z_2,t_2)) = E(V_0(z_1,t_1)V_0(z_2,t_2)),
~~\forall t_1, t_2,~~ \forall z_1,z_2,~~0 \leq z_1, z_2 \leq L
\label{eqn:potentialsecond}
\end{equation}
were the expectations of $V_0$ terms are taken over the joint
distributions of the inputs to all channels.

\section{Channel Model}
\label{sec:ChannelModel} In this section we obtain a stochastic
channel model based on the equation of motion obtained in
section~\ref{sec:propagation}. This stochastic model is simplified
to result in a channel model which incorporates the effects of XPM
which results in a multiplicative phase noise term.

\subsection{Effects of XPM}
Equation~(\ref{eqn:propagation}) describes the effect of XPM on
the relation between the input and output signal of the central
channel. Based on this equation the channel model or the
input-output relation can be described using the Green's function
$G(L;t,t')$, also known as the propagator~\cite{Feynman},
\begin{equation}
y(t)=x_{0}(L;t)=\int G(L;t,t')x_{0}(0;t')dt'+n(t), \label{eqn:inout}
\end{equation}
where $L$ is the length of the fiber and $n(t)=\int_{0}^{L}
n(z,t)dz$ takes into account the effect of all additive noise, which
we shall approximate with an additive white Gaussian random process,
independent of all inputs to the channel~\cite{Mitra}.

The Green's function is a function of the random potential $\nu
(z,t)$ and can be written in the form of a path
integral~\cite{Feynman},
\begin{equation}
G(L;t,t')=\int \mathcal{D}t(z)\exp \left[-\int_{0}^{L}
\frac{i}{2\beta_{2}} (\partial_{z}t(z))^{2}dz - i\int_{0}^{L} \nu
(t(z),z)dz \right],\label{eqn:G1}
\end{equation}
where the paths $t(z)$ start at $t'$, at $z = 0$ and end at $t$,
at $z = L$. This expression is encountered in the context of
Quantum Mechanics, where the roles of time and space are
interchanged. We shall simplify this expression to obtain a
channel model.
\begin{theorem}
\label{thm:channel:model} Simplifying Equation~(\ref{eqn:G1}),
combined with Equation~(\ref{eqn:inout}) result in the following
simplified channel model,
\begin{equation}
y(t) = \int \exp(-i\frac{(t-t')^{2}}{2\beta L}) \exp(-i L U(L;t',t))
x(t')dt' +  n(t)\label{eqn:model1},
\end{equation}
where
\begin{equation}
U(L;t',t) \sim \mathcal{N}
(0,\sigma^{2}_{U}(t-t')).\label{Udistribution}
\end{equation}
Here,
\begin{equation*}
\sigma^{2}_{U} = \frac{2P^{2}}{\beta_2 (\delta \nu)^2}
\sum_{n=1}^{N/2} {1 \over n},
\end{equation*}
where $P$ is the power in each channel, $\beta_2$ is the propagation
constant of the central channel, $\delta \nu$ is the channel
frequency spacing and $N$ is the number of channels.
\end{theorem}
\begin{proof}
The rest of section~\ref{sec:ChannelModel} is devoted to proving
Theorem~(\ref{thm:channel:model}). This is done by simplifying and
approximating the Greens function given by
Equation~(\ref{eqn:G1}).

The potential can be written as $\nu(z,t) = E(\nu(z,t)) + \delta
\nu(z,t)$. The average value of the random potential causes a
deterministic constant phase factor, which has no effect on the
information capacity of the channel. However, the fluctuations of
the random potential about its average value effect the phase of the
Green's function and hence the capacity of the chanel. Since the
average potential has no effect on our results we shall assume
$E(\nu(z,t)) = 0$, hence $\nu(z,t) = \delta \nu(z,t)$.

\subsection{``Green's Function" Approximation}
\label{Green} In this section we shall approximate the Green's
function given by Equation~(\ref{eqn:G1}) using stochastic and
physical properties of the random potential.

Refereing to~\cite{Feynman} we could approximate the expression for
$G(L;t,t')$ by dividing both space and time into small intervals. We
divide the time interval $(t',t)$ into $M$ equal subintervals,
$\Delta t$, resulting in $\{t_i\}_{i=0}^{M}$, where $t_0 = t'$ and
$t_M = t$. Also, we shall divide the fiber length into $M$ equal
subintervals $\Delta L$.

In all that follows we shall be concerned with the time interval
$(t',t)$, i.e. $\nu(z,t")=0$ if $t" \notin (t',t)$. However, we
shall assume that the inputs start from a distant past and will
continue into the distant future. This assumption is justified
since the signals in different channels are not necessarily
synchronized. And so we have~\cite{Feynman},
\begin{eqnarray*}
\int \mathcal{D}t(z)&\approx& \int \int ... \int dt_{M-1} dt_{M-2}
... dt_{1},\\
\int_{0}^{L} \frac{i}{2\beta_{2}}(\frac{\partial t(z)}{\partial
z})^{2}dz &\approx& \sum_{0}^{M}
\frac{i}{2\beta_{2}}[\frac{t_{k}-t_{k-1}}{\Delta L}]^{2}\Delta L,\\
\int_{0}^{L}\nu (t(z),z)dz &\approx&\sum_{0}^{M}\nu (t_{k})\Delta L.
\end{eqnarray*}
These approximations result in,
\begin{eqnarray}
G(L;t,t')&=&\int \mathcal{D}t(z) \exp\left[-\int_{0}^{L}
\frac{i}{2 \beta_{2}} (\partial_{z}t(z))^{2}dz - i \int_{0}^{L}
\nu (t(z),z)dz \right] {\nonumber} \\
&\approx&\int \int ... \int \exp \left[ -\sum_{k=1}^{M-1}
\frac{i}{2\beta_{2}} [\frac{t_{k}-t_{k-1}}{\Delta L}]^{2}\Delta L
-i\sum_{k=1}^{M-1} \nu (t_{k}) \Delta L \right] dt_{M-1}dt_{M-2} ...
dt_{1} {\nonumber} \\
&=& \int \int ... \int \prod_{k=1}^{M-1} \exp \left(
-\frac{i}{2\beta_{2}} \frac{(t_{k}-t_{k-1})^{2}}{\Delta L}\right)
\exp \left(- i \nu (t_{k}) \Delta L \right) dt_{M-1}dt_{M-2} ...
dt_{1}.\nonumber\\
&&\label{eqn:randompotential}
\end{eqnarray}
Consider the terms containing the random potential in
Equation~(\ref{eqn:randompotential}) given by,
\begin{equation*}
\exp(- i \nu (t_{k}) \Delta L) = 1 - i \nu (t_{k}) \Delta L - (\nu
(t_{k}) \Delta L)^{2} + ... = 1 - i \nu (t_{k}) \Delta L + O((\Delta
L)^2),
\end{equation*}
which can be approximated by neglecting the $O(\Delta L)^2$ terms.
This approximation is justified considering nominal values for the
parameters, see~\cite{Agrawal}. Hence,
\begin{eqnarray*}
G(L;t,t') &\approx& \int \int ... \int \prod_{k=1}^{M-1} \exp \left(
-\frac{i}{2\beta_{2}} \frac{(t_{k}-t_{k-1})^{2}}{\Delta L} \right)
\left(1 - i \nu (t_{k}) \Delta L \right) \, dt_{M-1}dt_{M-2} ...
dt_{1}
\end{eqnarray*}
evaluating these integrations results in,
\begin{eqnarray}
G(L;t,t') &\approx&
\exp\left(-\frac{i(t-t')^{2}}{2\beta_{2}L}\right)\nonumber\\
&-& i L \int_{0}^{1} \int_{-\infty}^{\infty}
\exp\left(-\frac{i}{2\beta_{2}L}
[\frac{(t-t_{\alpha})^{2}}{\alpha} +
\frac{(t_{\alpha}-t')^{2}}{1-\alpha}]\right)\nu(t_{\alpha})
dt_{\alpha} d\alpha\nonumber\\
&=&\exp\left(-\frac{i(t-t')^{2}}{2\beta_{2}L}\right) - i L
U(L;t',t)\label{eqn:U}
\end{eqnarray}
Equation~(\ref{eqn:U}) consists of two terms, the first term is a
deterministic phase shift, while, the second term involves the
weighted integration of the random potential $\nu(t_{\alpha})$,
resulting in the random process $U(L;t,t')$. We need to characterize
the distribution of the random process $U(L;t,t')$.

\subsection{Distribution of $U(L;t,t')$}
In this section we shall characterize the distribution of
$U(L;t,t')$. This will result in characterizing the distribution
of our approximation to the Greens function.
\begin{lemma} \label{lemma:U} The distribution of the random process $U(L;t',t)$
is given by,
\begin{equation}
U(L;t',t) \sim \mathcal{N}
(0,\sigma^{2}_{\nu}(t-t')).\label{eqn:Udistribution}
\end{equation}
with
\begin{equation*}
\sigma^{2}_{\nu} = \frac{2P^{2}}{\beta_2 (\delta \nu)^2 }
\sum_{n=1}^{N/2} {1 \over n} \approx \frac{2P^{2}}{\beta_2 (\delta
\nu)^2} \log(N/2)
\end{equation*}
\end{lemma}
\begin{proof}
An outline of the proof of Lemma~\ref{lemma:U} is given in the
appendix.
\end{proof}
In our derivation of the distribution of the Greens function we
neglected the higher order terms in $\Delta L$. If we follow the
same tedious calculations for higher orders of $\Delta L$ we get
the following approximation for the Greens function,
\begin{eqnarray}
G(L;t,t') &\approx& \exp\left(-i\frac{(t-t')^{2}}{2\beta
L}\right)(1-i L U(L;t',t) + {(-i)^2 L^2 U^{2}(L;t',t) \over 2}+ \cdots)\nonumber\\
&=& \exp\left(-i\frac{(t-t')^{2}}{2\beta L}\right) \exp(-i L
U(L;t',t)) \label{eqn:G}
\end{eqnarray}
where the distribution of $U(L;t',t)$ is given by
Equation~(\ref{eqn:Udistribution}). The proof of
Equation~(\ref{eqn:G}) follows from the proof of
Lemma~\ref{lemma:U} in the appendix and therefore not presented
here. Combining Equation~(\ref{eqn:G}) with
Equation~(\ref{eqn:inout}) results in the following approximate
channel model,
\begin{equation*}
y(t) = \int \exp(-i\frac{(t-t')^{2}}{2\beta L}) \exp(-i L U(L;t',t))
x(t')dt' +  n(t),
\end{equation*}
which completes the proof of Theorem~\ref{thm:channel:model}.
\end{proof}

\section{Simplified Channel Model}\label{sec:simplify}
In this section we use some physical characteristics of an optical
fiber to further simplify the channel model given by
Theorem~\ref{thm:channel:model}. Equation~(\ref{eqn:model1}) is a
standard representation of the following input output relation,
\begin{equation}
x_0(L;t) = \int \exp(-i\frac{(t-t')^{2}}{2\beta L}) \exp(-i L
U(L;t',t)) x_0(0;t')dt' +  n(t).\label{eqn:model2}
\end{equation}
As shown by Equation~(\ref{eqn:model2}), the output at time $t$ is
affected by the input at time $t'$. The propagation (travel) time of
the signal in the central channel is given by $\frac{L}{\nu_g}$,
where $\nu_g$ is the group velocity of the central channel,
see~\cite{Agrawal} and $L$ is the fiber length. Hence, the input
time interval $t'$, that affects the output at time $t$ is a window
in time around $t - \frac{L}{\nu_g}$.

Propagation along the fiber is dispersive, i.e. signals tend to
spread as they propagate along the fiber. However, from practical
considerations, we know that the propagation distances used in
practice are small enough to ensure that the signal spread is small
compared to the propagation delay $\frac{L}{\nu_g}$. Hence the input
time interval $t'$ that affects the output at time $t$ is small
compared to $\frac{L}{\nu_g}$.

As an example assume that the signals are bit streams of time
duration $T \approx {1 \over \delta \nu} = 20ps$. Nominal value
for the fiber parameters are given by, group velocity $\nu_g =
200,000 km/s$, propagation constant $\beta_2 = 20 {ps^2 \over km}$
and fiber length $L = 50 km$. In most practical situations no more
than $m = \pm 100$ neighboring bits will affect each bit. Hence,
we have $m T \approx 4 ps$ and $\frac{L}{\nu_g} \approx .25ms$ and
so the time interval that affects the output at each time is much
smaller than the propagation time of the signal. Hence,
\begin{equation}
t-t' \approx \frac{L}{\nu_g}.\label{eqn:velocity}
\end{equation}
Based on this approximation $U(L;t',t)$ is a stationary random
process in time, hence we'll drop the time dependance in
$U(L;t',t)$. We also absorb the $L$ in the channel model preceding
$U(L)$, into it's variance, resulting in,
\begin{equation*}
U(L) \sim \mathcal{N} (0,\sigma_{U}^{2} ({L \over \nu_g}) (L^2)).
\end{equation*}
The result of the approximation in~(\ref{eqn:velocity}) is to
``lump" the effect of a ``continuously injected multiplicative
noise" $U(L;t,t')$, along the fiber into a lumped multiplicative
noise $U(L)$, in the form of a random phase shift, at the end of
the fiber length $L$. The resulting channel model is given by,
\begin{eqnarray*}
y(t) &=& \int \exp(-i\frac{(t-t')^{2}}{2\beta L}) \exp(-i U(L))
x(t')dt' +  n(t)\\
&=& \exp(-i U(L)) \int \exp(-i\frac{(t-t')^{2}}{2\beta L})
x(t')dt' +  n(t)
\end{eqnarray*}
Note that $\int \exp(-i\frac{(t-t')^{2}}{2\beta L}) x(t')dt'$ is
the convolution of the input signal $x(t)$ with
$\exp(-i\frac{t^{2}}{2\beta L})$, which results in a deterministic
phase shift in the frequency domain which can be compensated for
and so it has no effect on the information capacity of the channel
or the capacity achieving input signal. As a result the input
signal can be represented as,
\begin{equation*}
\tilde{x}(t) = \int \exp(-i\frac{(t-t')^{2}}{2\beta L}) x(t')dt'.
\end{equation*}
Finally we have the following simplified channel model,
\begin{equation}
y(t) = \exp(-i U(L)) \tilde{x}(t) + n(t),\label{eqn:channelmodel}
\end{equation}
where $U(L) \sim \mathcal{N} (0,\sigma_{U}^{2} ({L \over
\nu_g})(L^2))$.

\section{channel capacity}\label{sec:capacity}
In this section we shall provide an upper bound to the information
capacity of the channel given by Equation~(\ref{eqn:channelmodel})
for the high SNR regime.

Consider the channel given by Equation~(\ref{eqn:channelmodel}),
since each channel is bandlimited, the inputs to each channel are
also bandlimited, hence our continues time channel model is
equivalent to a discrete time channel model, see~\cite{Gallager},
\begin{equation}
y_k = \exp(-i u_k) \tilde{x}_k + n_k\label{eqn:Discreatmodel}
\end{equation}
where $u$ and $n$ are independent i.i.d sequences with $u_k \sim
\mathcal{N} (0,\sigma_{U}^{2} ({L \over \nu_g})(L^2))$ and $n_k \sim
\mathcal{N} (0,\sigma_{N}^{2})$.

Consider the channel given by Equation~(\ref{eqn:Discreatmodel}).
The input to the channel is given by $\{x_k\}$ with $x_k \in
\mathbb{C}$ with an input power constraint $E(\frac{1}{n}
\sum_{k=1}^{n} |x_k|^2) \leq P$. The additive noise $\{n_k\}$ is
an i.i.d. sequence of circularly symmetric complex Gaussian random
variables with $n_k \sim \mathcal{N} (0,\sigma_{N}^{2})$, also the
phase noise $\{u_k\}$ is an i.i.d. sequence of Gaussian random
variables with $u_k \sim \mathcal{N} (0,\sigma_{U}^{2})$, with the
distribution of $\{u_k\}$ being independent of the input power.
Also $u_k$ has finite entropy. Note that all three process
$\{x_k\}$, $\{n_k\}$ and $\{u_k\}$ are mutually independent. Given
this setting we have the following lemma,

\begin{lemma} Given the above setting, the capacity of the channel given by
Equation~(\ref{eqn:Discreatmodel}) is upper bounded by,
\label{lemma:capacity}
\begin{eqnarray}
C &\leq&
\frac{1}{2}\log\left(1+2\pi^2e^{-2h(u)}\frac{P}{\sigma_N^2}\right)+o(1)\nonumber\\
&=& \frac{1}{2}\log\left(1+(\frac{1}{P\sigma_N^2})(\frac{\pi}{e})
(\frac{\beta_2 (\delta \nu)^2}{2\log(N/2)})
(\frac{\nu_g}{L^3})\right)+o(1),
\end{eqnarray}
in the high SNR regime. Where the $o(1)$ term tends to zero as
$\frac{P}{\sigma^2_N}$ tends to infinity.
\end{lemma}
The proof of Lemma~\ref{lemma:capacity} follows from the proof
in~\cite{Lapidoth}. As an example, consider the following nominal
values for a typical WDM optical fiber, $\beta_2=20
\frac{ps^2}{km}$, $\gamma=1.2 (W km)^{-1}$, $\delta \nu = 50GHz$,
$\nu_g=200,000\frac{km}{s}$, $L=50km$, and $N=100$. which result
in the capacity upper bounded by,
\begin{eqnarray*}
C &\leq& \frac{1}{2}\log\left(1+\frac{.0118}{P\sigma_N^2}\right)
\end{eqnarray*}

\section{Conclusion}

An upper bound to the information capacity of a wavelength-division
multiplexed optical fiber communication system is derived in a model
incorporating the nonlinear propagation effects of cross-phase
modulation (XPM). We modeled the effects of the continuously
injected phase noise due to the cross phase modulation nonlinearity
as a lumped multiplicative phase noise at the end of the fiber. This
model leads to an upper bound to the capacity of a WDM fiber optical
communication channel in the high SNR regime. This upper bound is in
agreement with the lower bound derived by Mitra et al.~\cite{Mitra}.

Future directions include better models for the effect of various
fiber nonlinearities, including XPM. Also upper bounds for the low
SNR regime and tighter upper bounds for the high SNR regime could be
derived as we believe our upper bound could be improved.

\section*{Acknowledgment}
The auther would like to thank Professor A.~El~Gamal for his
guidance through the work, also Dr.~P.~Mitra for many valuable
discussions.

\section {Appendix}
\textbf{Outline of proof of Lemma~\ref{lemma:U}} \label{app:U}\\
Carrying out the tedious calculations,
\begin{eqnarray*}
U(L;t,t') &=& \int_{0}^{1} \int_{-\infty}^{\infty}
\exp\left(-\frac{i}{2\beta_{2}L}
\left[\frac{(t-t_{\alpha})^{2}}{\alpha} +
\frac{(t_{\alpha}-t')^{2}}{1-\alpha}\right]\right)\nu(t_{\alpha})
dt_{\alpha} d\alpha\\
&=&  \exp\left(\frac{(-i)(t - t')^{2}}{2\beta_{2} L}\right)
\int_{0}^{1} U_{\alpha}(L;t',t) d\alpha,
\end{eqnarray*}
as mentioned in Section~\ref{Green}, $\nu(t_{\alpha})$ is zero
outside the time interval $(t',t)$, hence,
\begin{eqnarray*}
U_{\alpha}(L;t',t) &=& \int_{t'}^{t} \exp\left(-\frac{i}{2\beta_{2}
L \alpha (1-\alpha)} [t_{\alpha}-t(1-\alpha) -
t'(\alpha)]^{2}\right) \nu(t_{\alpha})
dt_{\alpha} d\alpha\\
&=& \int_{-(1-\alpha)(t-t')}^{\alpha(t-t')}
\exp\left(-\frac{i}{2\beta_{2} L \alpha
(1-\alpha)}(t_{\alpha})^{2}\right) \nu \left(t_{\alpha}+
t(1-\alpha) + t'(\alpha)\right) dt_{\alpha} d\alpha\\
&=_{D}& \int_{-(1-\alpha)(t-t')}^{\alpha(t-t')} \exp\left[-\frac{i}
{2\beta_{2} L \alpha (1-\alpha)}
(t_{\alpha})^{2}\right]\nu(t_{\alpha}) dt_{\alpha}
\end{eqnarray*}
where $=_{D}$ equates the distribution of both sides, which follows
from the stationarity of $\nu(t)$ and the fact that both
$t_{\alpha}+ t(1-\alpha) + t'(\alpha)$ and $t_{\alpha}$ cover the
integration interval $(-(1-\alpha)(t-t'), \alpha(t-t'))$, for any
$\alpha \in (0,1)$. Note that we are only interested in the
distribution of $U_{\alpha}(L;t',t)$.

Since $\nu(t)$ is a Gaussian process, we have that
$U_{\alpha}(L;t',t)$ is also a Gaussian random process~\cite{
Kailath}, i.e. $U_{\alpha}(L;t',t) \sim \cal{N}(\mu,\sigma)$,
where it can be can shown,
\begin{eqnarray*}
\mu &=& E\left[\int_{-(1-\alpha)(t-t')}^{\alpha(t-t')}
\exp\left(-\frac{i} {2\beta_{2} L \alpha (1-\alpha)}
(t_{\alpha})^{2}\right)\nu(t_{\alpha}) dt_{\alpha}\right]=0,\\
\sigma^{2} &=& E [\int \exp\left(-\frac{i}{2\beta_{2} L \alpha
(1-\alpha)} (t_{1})^{2}\right)\nu(t_{1}) dt_{1} \int
\exp\left(+\frac{i} {2\beta_{2} L \alpha
(1-\alpha)}(t_{2})^{2}\right)\nu(t_{2}) dt_{2}]\\
&=& \sigma^{2}_{\nu}(t-t').
\end{eqnarray*}
Hence $U_{\alpha}(L;t',t) \sim
\mathcal{N}(0,\sigma_{\nu}^{2}(t-t'))$, where $\sigma^{2}_{\nu} =
E[\nu(t)^2]$ is the power of the Gaussian process $\nu(t)$ given
by,
\begin{equation*}
\sigma^{2}_{\nu} = \frac{2P^{2}}{\beta_2 (\delta \nu)^2}
\sum_{n=1}^{N/2} {1 \over n} \approx \frac{2P^{2}}{\beta_2 (\delta
\nu)^2} \log(N/2),
\end{equation*}
for details of this derivation see~\cite{Mitra}.

Note that $U_{\alpha}(L;t',t)$ is a jointly Gaussian random
process in $\alpha$. Hence, integrating $U_{\alpha}(L;t',t)$
results in a Gaussian random process~\cite{Kailath}, i.e. $
U(L;t',t) \sim \mathcal{N} (\mu_{U},\sigma_{U})$, where it can be
shown,
\begin{eqnarray*}
\mu_U &=& E [\int_{0}^{1} U_{\alpha}(L;t',t) d\alpha] = 0,\\
\sigma^{2}_U &=& E [\int_{0}^{1} U_{\alpha}(L;t',t) d\alpha
\int_{0}^{1} U^{*}_{\alpha'}(L;t',t) d\alpha']\\
&=& \sigma^{2}_{\nu} (t-t'). \label{eqn:sigmaU}
\end{eqnarray*}
And so,
\begin{equation*}
U(L;t',t) \sim \mathcal{N}
(0,\sigma^{2}_{\nu}(t-t')).\label{Udistribution}
\end{equation*}

\end{document}